\documentclass[floatfix,amssymb,prl,twocolumn,superscriptaddress,nofootinbib, aps]{revtex4-1}
		
\usepackage{amssymb,amsmath,verbatim,mathtools,needspace,enumitem,etoolbox,graphicx,physics,microtype,afterpage,bm,soul}
\usepackage[dvipsnames]{xcolor}
\definecolor{linkcolor}{rgb}{0.0,0.3,0.5}
\usepackage[unicode, colorlinks=true, linkcolor=linkcolor, citecolor=linkcolor, filecolor=linkcolor,urlcolor=linkcolor, pdfusetitle]{hyperref}
\usepackage{orcidlink}
\usepackage[all]{hypcap}
\usepackage[T1]{fontenc}
\usepackage[utf8]{inputenc}
\usepackage{tabularx}
\usepackage{float}
\usepackage{lmodern}
\usepackage{ragged2e}
\allowdisplaybreaks
\usepackage{tikz}
\usepackage{color}
\usepackage{framed}
\usepackage{hyperref}
\hypersetup{colorlinks, citecolor=darkblue, linkcolor=black, urlcolor=darkblue}
\definecolor{rossos}{cmyk}{0,1,1,0.55}
\definecolor{bluscuro}{rgb}{0.15, 0.2, .85}
\definecolor{bluchiaro}{cmyk}{1,.3,0.,0.1}
\definecolor{ForestGreen}{rgb}{0.13, 0.55, 0.13}
\definecolor{darkblue}{rgb}{0,0, 1.39}

\newcommand{\be}{\begin{equation}}
\newcommand{\ee}{\end{equation}}

\newcommand{\llp}{\left [}
\newcommand{\rrp}{\right ]}
\newcommand{\lp}{\left (}
\newcommand{\rp}{\right )}

\newcommand{\bmr}{{\bm r}}
\newcommand{\zsl}{z_{\rm sl}}
\newcommand{\zlo}{z_{\rm lo}}
\newcommand{\zso}{z_{\rm so}}
\newcommand{\bo}{{\bm b}_{\rm o}}
\newcommand{\bl}{{\bm b}_{\rm l}}
\newcommand{\bb}{{\bm b}}

\def\lsim{\mathrel{\rlap{\lower4pt\hbox{\hskip0.5pt$\sim$}}
    \raise1pt\hbox{$<$}}}         
\def\gsim{\mathrel{\rlap{\lower4pt\hbox{\hskip0.5pt$\sim$}}
    \raise1pt\hbox{$>$}}}         

\newcommand{\jhu}{William H.\ Miller III Department of Physics and Astronomy, Johns Hopkins University, \\ 3400 North Charles Street, Baltimore, Maryland, 21218, USA}
\newcommand{\UPenn}{Center for Particle Cosmology, Department of Physics and Astronomy, University of Pennsylvania 209 South 33rd Street, Philadelphia, Pennsylvania 19104, USA}

\begin{document}

\title{Scattering perspective on gravitational lensing}

\author{Mariana Carrillo Gonzalez\orcidlink{0000-0003-1119-9097}}
\email{m.carrillo-gonzalez@imperial.ac.uk \\ m.carrillo-gonzalez@soton.ac.uk}
\affiliation{Abdus Salam Centre for Theoretical Physics, Imperial College, London, SW7 2AZ, United Kingdom}
\affiliation{School of Physics and Astronomy, University of Southampton,
Highfield, Southampton, SO17 1BJ, United Kingdom}

\author{Valerio De Luca\orcidlink{0000-0002-1444-5372}}
\email{vdeluca2@jh.edu}
\affiliation{\jhu}

\author{Alice Garoffolo\orcidlink{0000-0002-0125-4577}}
\email{aligaro@sas.upenn.edu}
\affiliation{\UPenn}

\author{Julio Parra-Martinez\orcidlink{0000-0003-0178-1569}}
\email{julio@ihes.fr}
\affiliation{Institut des Hautes \'Etudes Scientifiques, 91440 Bures-sur-Yvette, France}

\author{Mark Trodden\orcidlink{0000-0001-6864-6711}}
\email{trodden@upenn.edu}
\affiliation{\UPenn}


\begin{abstract}
\medskip
\noindent
Gravitational waves propagating across gravitational potentials undergo lensing effects that, in the wave-optics regime, manifest as frequency-dependent amplitude and phase modulations. In this work, we revisit the diffraction integral formalism of gravitational lensing and demonstrate that it admits a natural and transparent interpretation within the framework of scattering theory. We establish a direct correspondence between the lensing amplification factor and the scattering amplitude of waves propagating in curved spacetime, clarifying how familiar lensing limits map onto distinct scattering regimes. In particular, we show that the  diffraction integral  matches exactly the eikonal limit of the scattering amplitude at lowest post-Minkowskian order, after a change in coordinates and the inclusion of finite-distance effects. We further extend the standard formalism by including subleading corrections to the post-Minkowskian and eikonal approximations. Our results provide a unified theoretical framework for the interpretation of lensed gravitational-wave signals and open the way to more accurate waveform modeling for future lensed observations.

\end{abstract}
\maketitle

\vspace{0.1cm}
\noindent{{\bf{\em Introduction.}}} 
The advent of gravitational-wave (GW) astronomy has enabled direct studies of compact objects and their populations across cosmic time. Since the first detection of a binary black hole merger in 2015~\cite{LIGOScientific:2016aoc}, the LIGO-Virgo-KAGRA Collaboration has progressively expanded its sensitivity and frequency coverage~\cite{LIGOScientific:2018mvr, LIGOScientific:2020ibl, KAGRA:2021vkt}, recently releasing its fourth observing run catalog~\cite{LIGOScientific:2025hdt,LIGOScientific:2025slb}.
Together with the three preceding observing runs, the full catalog now includes hundreds of confirmed compact-binary coalescences, spanning a wide range of source masses, mass ratios, and redshifts. As the number of detections grows, especially with the advent of next generation detectors such as the Einstein Telescope, Cosmic Explorer and LISA~\cite{ET:2025xjr, Branchesi:2023mws,LISA:2024hlh}, the probability that some of these signals have been affected by gravitational lensing also increases, making wave propagation effects through intervening structures an important ingredient in the interpretation of current and future GW data~\cite{Hannuksela:2019kle,Liu:2020par,LIGOScientific:2021izm,Jung:2017flg, Wempe:2022zlk, LIGOScientific:2023bwz,Oguri:2018muv, Yin:2023kzr, Sereno:2010dr, Lai:2018rto,Urrutia:2021qak,Baker:2016reh}. Intriguingly, the recently reported event GW231123 has been highlighted as a potential candidate for providing observational evidence of GW lensing~\cite{LIGOScientific:2025rsn}.

Depending on the ratio between the GW wavelength and the characteristic scale of the lens potential, gravitational lensing operates in qualitatively different regimes. In the geometric-optics limit, where the wavelength is negligible compared to the lens scale, the signal is described as a sum of distinct images associated with stationary paths of the Fermat potential, each characterized by a magnification and a time delay~\cite{Schneider:1992bmb, Meneghetti:2020yif}. In contrast, when the wavelength becomes comparable to the Schwarzschild radius of the lens, wave effects become dominant and diffraction and interference imprint frequency-dependent modulations on the observed waveform~\cite{Ohanian:1974ys, Bliokh, Bontz:1981rvr, Deguchi:1986zz,Takahashi:2003ix,Schneider:1992bmb, Nakamura:1999uwi} (for a recent review on gravitational lensing in the context of GWs, see Ref.~\cite{Grespan:2023cpa}).

A general treatment of GW lensing in the wave regime is provided by the diffraction integral, which captures the coherent superposition of all propagation paths through the lens plane~\cite{Takahashi:2003ix, Oguri:2020ldf, Nakamura:1999uwi, Takahashi:2004mc, Takahashi:2005sxa, Takahashi:2005ug}.
This formulation contains both the geometric and diffractive limits as special cases: in the high-frequency limit, the integral reduces to a discrete sum over stationary paths corresponding to the standard image positions, while in the low-frequency limit it describes a single, smooth amplification factor accounting for interference and phase modulation~\cite{Jow:2022pux}. The diffraction integral thus provides a unified mathematical description of GW lensing that naturally bridges the wave and geometric regimes~\cite{Takahashi:2003ix, Oguri:2020ldf, Nakamura:1999uwi, Takahashi:2004mc, Takahashi:2005sxa, Takahashi:2005ug}.

Given the upcoming observational prospects, the literature focusing on wave-optics effects has been growing. Several groups have developed numerical codes to compute the diffraction integral efficiently~\cite{yeung2024wolensing, Villarrubia-Rojo:2024xcj,Feldbrugge:2019fjs,Diego:2019lcd}. 
These codes have been used to study the properties of different lens models~\cite{Tambalo:2022wlm,Caliskan:2022hbu}, including rotating lenses~\cite{Bonga:2024orc,Kubota:2023dlz,Kubota:2023dlz,Li:2022izh}, lenses in binary systems~\cite{Feldbrugge:2020ycp,Mehrabi:2012dy} or the intriguing possibility that the lens is given by a different dark matter model~\cite{Singh:2025uvp}. 
Wave-optics effects also play an important role in placing constraints on the abundance of primordial black holes~\cite{Sugiyama:2019dgt,Diego:2019rzc}.
With a particular focus on the space-based interferometer LISA, probabilities of observing such effects and parameter estimations for the lens properties can be found in~\cite{Caliskan:2023zqm,Caliskan:2022hbu}.

Despite its broad applicability, the physical meaning of the diffraction integral beyond its computational use has remained somewhat obscure. In particular, it is difficult to maintain full control over the approximations on which it relies, both to ensure that predictions are not extrapolated beyond their domains of validity and to develop systematic improvements.

In this paper, we address this issue by connecting the diffraction integral description of lensing to the more general framework of scattering theory. Although the analogy between gravitational lensing and wave scattering has long been recognized, the formal connection between the diffraction integral formalism and the general theory of wave propagation—such as that provided by scattering theory~\cite{Newton:1982qc}—has not been systematically explored. Yet, the structure of a lensing event closely mirrors that of a scattering process, in which an incoming wave interacts with an external potential and emerges with a modified amplitude and phase~\cite{Futterman:1988ni, Vishveshwara:1970zz, Press:1971wr, Macquart:2004sh}. This perspective suggests that gravitational lensing can be understood as the scattering of gravitational waves by the gravitational potential of the lens, offering a conceptually transparent and physically unified framework with well-defined expansion parameters governing the validity of the results~\cite{Virbhadra:1999nm, Bozza:2001xd, Bozza:2002af, Pijnenburg:2024btj, Chan:2025wgz, Yin:2023kzr, Motohashi:2021zyv}. More specifically, gravitational lensing corresponds to the gravitational Compton scattering process, which has largely been studied in the on-shell program for gravitational waves from binary systems \cite{Bjerrum-Bohr:2014zsa,Bjerrum-Bohr:2016hpa,Bai:2016ivl,Chi:2019owc,AccettulliHuber:2020oou,Bastianelli:2021nbs,Chen:2022clh,Comberiati:2024uuc,Cangemi:2023bpe, Ivanov:2024sds, Correia:2024jgr, Caron-Huot:2025tlq, Bautista:2021inx, Bautista:2021wfy, Bautista:2022wjf,Bjerrum-Bohr:2025bqg, Akpinar:2025byi, DiVecchia:2023frv, KoemansCollado:2019ggb, Aoude:2022thd, Chen:2021kxt,Chiodaroli:2021eug,Aoude:2022trd,Bjerrum-Bohr:2023jau,Bjerrum-Bohr:2023iey,Vazquez-Holm:2025ztz}.

In this work, we develop a scattering-based interpretation of GW lensing and establish a direct correspondence between the lensing amplification factor and the scattering amplitude of waves interacting with a lens. 
This mapping clarifies how familiar lensing limits translate into distinct scattering regimes, showing in particular that the diffraction integral corresponds to the eikonal limit of the scattering amplitude.

The paper is organized as follows. We begin by reviewing the frameworks of the diffraction integral and of scattering amplitudes. We then demonstrate how these two approaches can be mapped onto each other, both in the Born regime and via partial-wave expansions. Finally, we discuss post-Minkowskian and beyond-eikonal corrections, providing extensions to the standard diffraction-integral results. Throughout, we adopt natural units with $\hbar = c = 1$, and use boldface symbols to denote vectors in either two or three dimensions.

\begin{figure}[t!]
    \centering
\includegraphics[width=0.42\textwidth]{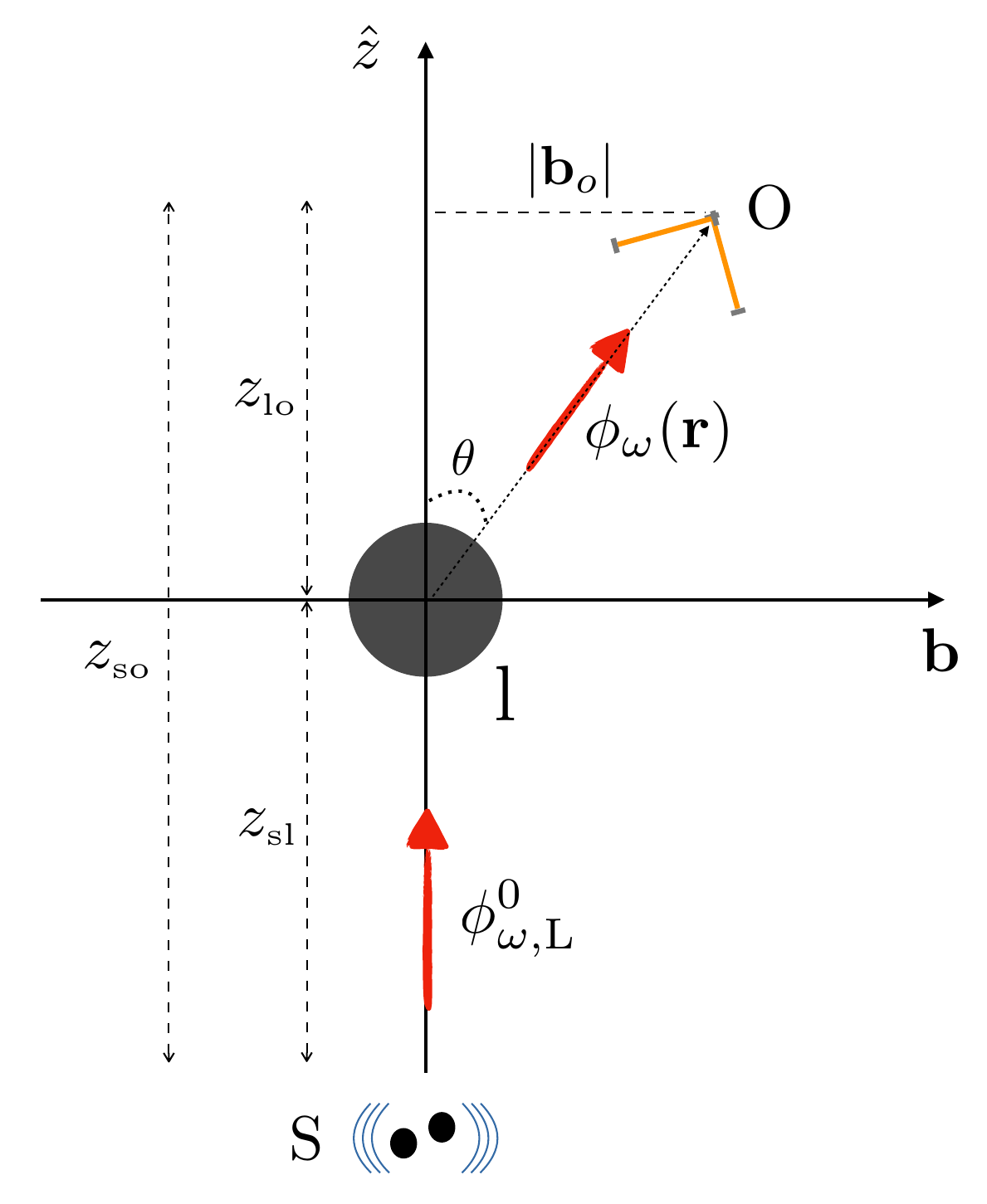}
    \caption{Schematic illustration of the lensing/scattering event, showing an incoming scalar wave $\phi^0_{\omega, {\rm L}}$ interacting with a gravitational lens (placed at the center of the reference frame). The interaction generates an outgoing wave $\phi_{\omega}$ that eventually reaches an observer located far away.
    }
\label{fig:setup}
\end{figure}

\vspace{0.1cm}
\noindent{{\bf{\em The lensing diffraction integral.}}} 
To set the stage for the derivation of the lensing diffraction integral, we refer to the configuration shown in Fig.~\ref{fig:setup}. The lens $(\rm l)$ is placed at the origin, the source $(\rm s)$ of the waves lies at $\bmr_{\rm s} = (0, 0, -\zsl)$, while the observation point $(\rm o)$ is located at $\bmr = (\bo, \zlo)$,  with $z_{ij}$ denoting the distances between the locations $i$ and $j$, and $\bo$ the transverse distance on the observer plane.\footnote{Let us mention that, in a cosmological setting, one should take into account the cosmic expansion in the relative distances.}
In the standard lensing framework, the lens is modeled as a Newtonian, static potential (later, we will relax this assumption to incorporate relativistic corrections). Despite the spin-2 nature of GWs, we will consider the simple case of scalar fields, hence phenomena such as the mixing of different polarization modes~\cite{Andersson:2020gsj, Oancea:2022szu} will be ignored. While these corrections might be relevant in the Born regime~\cite{Garoffolo:2022usx}, they are suppressed in the eikonal limit, which we adopt in many parts of this work (see the {\it Beyond eikonal corrections} Section for additional details). 
Therefore, the propagation equation reads~\cite{Grespan:2023cpa}
\be \label{eq:HelmholtzLensing}
\llp \nabla^2_\bmr + \omega^2 \lp 1 - 4 U_{\rm L} (\bmr) \rp \rrp \phi_{\omega} ( \bmr) = 0\,,
\ee 
up to first order in the gravitational potential of the lens, $U_{\rm L}$, and where we have performed a Fourier transformation to the frequency domain, with mode frequency $\omega$.  For distant GW sources, the emission can be approximated as pointlike, so that the unlensed signal is typically modeled as a spherical wave of the form $\phi^0_{\omega,{\rm L}} (\bmr) = e^{i\omega|{\bm r}-{\bm r}_{\rm s}|}/|{\bm r}-{\bm r}_{\rm s}|$. By introducing an {\it amplification factor} $F(\bmr )$ for the full lensed wave, 
\be\label{eq:DefAmplificationFactor}
\phi_{\omega} (\bmr) \equiv F(\bmr ) \, \times \, \phi^0_{\omega, {\rm L}} (\bmr)\,,
\ee 
one can rearrange Eq.~\eqref{eq:HelmholtzLensing} to obtain the form
\be\label{eq:EqF}
\partial_{\tilde r}^2 F + 2 i \omega \partial_{\tilde r} F + \frac{1}{{\tilde r}^2} \nabla^2_{\tilde {\bm \theta}} F = 4 \omega^2 U_{\rm L} F\,,
\ee 
written in spherical coordinates $\{ \tilde{\theta}, \tilde{\varphi} \}$ centered on the source, with $\tilde r = |\bmr - \bmr_{\rm s}|$ and $ \nabla^2_{\tilde {\bm \theta}}$ the Laplacian on the sphere. As shown in~\cite{Nakamura:1999uwi} (see also~\cite{Braga:2024pik, Feldbrugge:2019fjs} for alternative derivations), one solution of Eq.~\eqref{eq:EqF} can be found under the following set of assumptions, namely
\begin{align}
\label{eq:DiffIntAssumptions-para}
&\nabla^2_{\tilde {\bm \theta}} \approx \partial^2_{\theta} + {{\theta}}^{-1} \partial_{{\theta}} + {\theta}^{-2} \partial_{{\varphi}} \,, \\
\label{eq:DiffIntAssumptions-eik}
&|\partial_{\tilde r}^2 F| \ll |  \omega \partial_{\tilde r} F| \,.
\end{align} 
Eq.~\eqref{eq:DiffIntAssumptions-para} expresses the {\it paraxial} (or small-angle) approximation, in which the  wave  propagates predominantly along a single direction (taken here as the $z$-axis), such that it makes only small angles with that axis, $\tilde{\bm\theta} \ll 1$.\footnote{When expanding the  spherical Laplacian in the paraxial approximation, spherical coordinates reduce to cylindrical ones, so that centering the $z-$axis on the source or on the lens is equivalent. This also allows one to consider angles centered on the lens, $\{ \theta, \varphi \}$.}
On the other hand, Eq.~\eqref{eq:DiffIntAssumptions-eik} is typically identified as the {\it eikonal} approximation in the lensing literature, despite its practical interpretation often being ambiguous. As we show in this work, this assumption is equivalent to the conventional eikonal approximation used in scattering theory, corresponding to the regime of large impact parameters compared to the GW wavelength. These assumptions allow one to convert the wave equation into a Schr\"odinger-like form, which can be solved in terms of a path integral according to the expression~\cite{Nakamura:1999uwi}
\begin{align}
F({\bm r}) = \int \mathcal{D} \tilde{\bm \theta} (\tilde{r}') \exp \left\{ i \int_0^{\tilde{r}} {\rm d} \tilde{r}' \, \mathcal{L} [\tilde{r}', \tilde{\bm \theta}(\tilde{r}'),\dot{\tilde{\bm \theta}}( \tilde{r}')]\right\}\,,
\end{align}
where the classical Lagrangian is given by $\mathcal{L} [\tilde{r}', \tilde{\bm \theta},\dot{\tilde{\bm \theta}}] = \omega \left[ \frac{1}{2}\tilde{r}'^2 |\dot{\tilde{\bm \theta}}|^2  - 2 U_{\rm L}(\tilde{r}',\tilde{\bm \theta}) \right]$,
where the radial function  $\tilde{\bm \theta} (\tilde{r}')$ represents the path from the source to the observer at ${\bm r} = \tilde{{\bm r}} + {\bm r_s}$, $\dot{\tilde{\bm \theta}} = d \tilde{\bm \theta}/d \tilde{r}'$, and the integral is over all possible paths.

By further invoking the {\it thin-lens} approximation,  {\it i.e.} assuming that the lens extension along the optical axis is small compared to the relevant distances, one finds that, upon discretizing the path integral, all two-dimensional integrals except the one over the lens plane are Gaussian and can therefore be evaluated exactly. This is equivalent to stating that the paths contributing significantly to the phase integral are well approximated by a constant two-dimensional vector $\tilde{\bm \theta} (\tilde{r}) \simeq \bb/\zsl$ within the small region where $U_{\rm L}({\bm r}) \neq 0$, relative to the much larger scales $|\tilde{\bmr}|$ and $|\bm r_s|$, denoting with $\bb$ the transverse distance on the lens plane.  Under this approximation, the path integral thus reduces to a single two-dimensional integral over the lens plane, of the form~\cite{Nakamura:1999uwi,Grespan:2023cpa} 
\be\label{eq:DiffractionIntegral}
F_{\rm diff}(\bb_o) = \frac{\omega}{2\pi i } \frac{\zso}{ \zsl \zlo} \int \dd^2 \bb \, e^{i \omega t_d(\bb , \bo) - i \omega \psi (\bb)}\,.
\ee 
This expression is dubbed the {\it diffraction integral}, in which
\be\label{eq:TimeDelayGeom}
t_d(\bb, \bo)=\frac{\zso}{2 \zsl \zlo} \left(\bb -\frac{\zsl}{\zso} \bo\right)^2 \ ,
\ee
is the geometric time delay associated to the free wave propagation and 
\be \label{eq:2DPot}
\psi ({\bm b}) \equiv 2 \int \dd z \: U_{\rm L} \lp \bb,z \rp  \ ,
\ee 
is the gravitational time delay induced by the lens, which takes the form of a two-dimensional gravitational potential of the lens in its plane. 
Eq.~\eqref{eq:DiffractionIntegral} thus describes lensing as a sum over all possible scattering points on the lens plane, with each point weighted by an  action given by the sum of the geometric and gravitational time delays, {\it i.e.} the Fermat potential.

In the high-frequency limit, the integral in Eq.~\eqref{eq:DiffractionIntegral} is dominated by the saddle points of the total phase satisfying $\nabla_{\bb} [t_d (\bb, \bo) - \psi(\bb)] = 0$, thereby recovering the standard geometric-optics description, based on the lens equation
\be \label{eq:b_lens_obs}
\bb_{I} =\frac{\zsl}{\zso} \bo+\frac{\zsl\zlo}{\zso}\nabla_{\bb_I}\psi(\bb_I) \,,
\ee
where the label $I$ indicates the (potentially multiple) ``images'' associated to the stationary points of the phase. Then, the diffraction integral simplifies to the 
form~\cite{Nakamura:1999uwi, Grespan:2023cpa}
\be
\label{Fdiff-geome}
F_{\rm diff}(\bb_o) = \frac{\zso}{ \zsl \zlo} \sum_I \frac{e^{i \omega [t_d (\bb_I, \bo) - \psi(\bb_I)] - i \pi n_I/2}}{\sqrt{\text{det}\,H_{ij}(\bb_I)}}\,,
\ee 
where $H_{ij}(\bb)=\partial_i \partial_j [t_d (\bb, \bo) - \psi(\bb)]$ is the Hessian of the phase,  $n_I$ is the Morse
index, describing its number of negative eigenvalues at $\bb_I$, and the sum runs over all the phase critical points. In other words,  Eq.~\eqref{Fdiff-geome} shows that, in the geometric-optics limit, the amplification factor is described by a superposition of (de-)magnified images.  


\vspace{0.1cm}
\noindent{{\bf{\em The scattering amplitude perspective.}}} 
In scattering theory, the incoming wave is usually taken as a plane wave \be
\phi^0_{\omega, {\rm S}}(\bmr) = e^{i {\bm p} \cdot {\bmr}} =  e^{i \omega z} = e^{i \omega r \cos \theta}\,,
\ee
where ${\bm p} = (0,0,\omega)$, $\theta$ is the angle between the optical axis and the observer position, and $ r = |\bmr| $ is the radial distance to the scattering center. 
One can solve for the full wave using the free Green's function for Eq.~\eqref{eq:HelmholtzLensing} as 
\begin{align}
\phi_{\omega}({\bmr})&=\phi^0_{\omega, {\rm S}}(\bmr)-\frac{1}{4\pi}\int {\rm d}^3 {\bmr}' \frac{e^{i\omega |{\bf{r-r'}|}}}{|{ \bmr-\bmr'}|} \ 4 \omega^2 U_{\rm L}({\bm r}') \phi_{\omega}({\bm r}')\    \nonumber \\
&\simeq\phi^0_{\omega, {\rm S}}(\bmr)-\frac{e^{i\omega r}}{4\pi r}\int {\rm d}^3 {\bm r}'e^{-i\omega {\bm r}\cdot{\bm r}'/r}  \ 4 \omega^2 U_{\rm L}({\bm r}') \phi_{\omega}({\bm r}')\   \nonumber \\
&\equiv \phi^0_{\omega, {\rm S}}(\bmr)+f(\theta) \frac{e^{i \omega r}}{r} \  ,  \label{eq:ScatteringProblem}
\end{align}
where we assume an asymptotically distant observer by taking the limit $|{\bm r}|\gg|{\bm r}'|$ and define the \emph{scattering amplitude} $f(\theta)$, which, in this limit, depends only on $\theta\sim |\bo|/\zlo$. This quantity encodes the physical information about the scattering event and can be used to derive scattering cross-sections. The scattering amplitude can equivalently be defined by considering the incoming GW with 3-momentum ${\bm p}$ along the $z$-direction and
outgoing momentum given by ${\bm p}'=\omega\bmr/r$. In the large-distance limit, ${\bm p}' \approx (\omega\bo /\zlo,\omega)$, while the momentum transfer in the center-of-mass frame is
\be \label{eq:MomentumTransfer}
{\bm q} = {\bm p}' - {\bm p} = ( \omega \bo / \zlo, 0)\,. 
\ee 
Given this, one can define the amplitude as a function of these 3-momenta:
\be
f({\bm p},{\bm p}')=- \frac{1}{4\pi}\int {\rm d}^3 {\bm r}'e^{-i {\bm p}'\cdot{\bm r}'}  \ 4 \omega^2 U_{\rm L}({\bm r}') \phi_{|{\bm p }|}({\bm r}') \ .
\ee
For the lensing framework at hand, the scattering of a gravitational wave by a localized lens can be modeled as the Compton scattering of a graviton or, in this case, a massless scalar with a point particle, described in terms of an effective theory with field \cite{Cheung:2018wkq,Bern:2019crd,Bern:2019nnu} or worldline \cite{Goldberger:2004jt, Mogull:2020sak} degrees of freedom. Following this approach, the corresponding quantum field theory scattering amplitude $\mathcal{M}$ can be related in the classical limit to the scattering amplitude $f$ as\footnote{We adopt the normalization of \cite{Correia:2024jgr}, which ensures consistency between the Lippmann-Schwinger and the Schrödinger-like equation used in the lensing case.} 
\begin{equation}
f(\theta) = \frac{\mathcal{M} ({\bm p}, {\bm p}')}{8 \pi \sqrt{s}} \,,
\end{equation}
where $\sqrt{s}$ is the total energy in the center of mass frame.  

In this paper, we primarily focus on the case of a Newtonian point-mass lens---and its relativistic extension described by a Schwarzschild black hole---although the relations we derive with scattering amplitudes hold more generally. To obtain the classical contribution in a perturbative regime, we consider the hierarchies
\begin{align}\label{eq:ClassicalScattering}
\frac{\hbar}{M} \ll G M \ll  b, \quad
\frac{\hbar}{M} \ll \frac{1}{\omega} \lesssim b \,,
\end{align}
where $M$ is the lens mass, related to its gravitational potential $U_{\rm L}$ as discussed below, while $b = |\bb|$. Following these assumptions, the classical amplitude will be expressed as a series expansion in $GM/b$, encoding the classical non-linearities. 
The scattering amplitude and lens potential are then related by the Lippmann-Schwinger equation~\cite{Iwasaki:1971vb,Cristofoli:2019neg} which, in momentum space, reads
\begin{align}\label{eq:LS}
\mathcal{M}({\bm p},{\bm p+ \bm q}) &= \mathcal{V}({\bm p},{\bm p+ \bm q}) \nonumber\\
&\quad + \frac{1}{2 \sqrt{s}} \int_{\bm k} \frac{\mathcal{M}({\bm p},{\bm k}) \mathcal{V}({\bm k},{\bm p+ \bm q})}{|{\bm k}|^2 - |{\bm p}|^2 - i \epsilon}\,.
\end{align}
Here, the incoming graviton and the lens have  four-momenta $p_1^\mu = (\omega, {\bm{p}})$ and $p_2^\mu = (\sqrt{\omega^2 + M^2}, -{\bm{p}})$, respectively,
such that $\sqrt{s} \simeq M$. We have also defined $\int_{\bm k} \equiv \int {\rm d}^3{\bm k}/(2\pi)^3$. Notice that the momentum-space potential $\mathcal{V}$ entering in the Lippmann-Schwinger equation is defined through the Fourier transform~\cite{Correia:2024jgr}
\be \label{eq:VFT}
\mathcal{V}(\bm{p}, \bm{p}+\bm{q}) 
= -2\sqrt{s} \int {\rm d}^3\mathbf{r'}\, e^{-i\bm{q}\cdot\mathbf{r'}}\, V(\mathbf{r'})\,,
\ee 
in terms of its coordinate-space counterpart $V(\mathbf{r})$, which in turn is related to that defined in Eq.~\eqref{eq:HelmholtzLensing} by $V (\bmr) = 4 \omega^2 U_{\rm L} (\bmr)$. 
Given the form of $V$, one can first determine $\mathcal{V}(\bm{p}, \bm{p}+\bm{q}) $, and then solve the Lippmann-Schwinger equation~\eqref{eq:LS} perturbatively.

\vspace{0.1cm}
\noindent{{\bf{\em Matching in the Born regime.}}} 
As a warm-up, we perform the matching between the diffraction integral and the scattering amplitude in the Born regime. 
This regime is described by the weak field limit, where $U_{\rm L} \ll 1$, and the observer position $\bmr$ is far away from the region of spacetime where the gravitational potential of the lens is non-negligible. In the following, we will also assume
\be\label{eq:ExpansionB}
 |\bo|\ll \zlo, \zsl \,, \qquad  |\bb|\ll (\zsl/\zso)|\bo|\,,
\ee
namely that the transverse distance on the observer plane is much smaller than the distances on the optical axis between lens-observer and lens-source, and that the region on the lens plane where the potential is non-negligible is smaller than the position of the observer on the same plane ({\it i.e.} $(\zsl/\zso)|\bo|$).

On the lensing side, the Born approximation is equivalent to expanding the exponential in Eq.~\eqref{eq:DiffractionIntegral} for small potentials $\psi$, yielding~\cite{Nakamura:1999uwi}
\begin{align}
F_{\rm diff}(\bmr) 
\approx 1 -  \frac{\omega^2 }{2\pi } \frac{\zso}{ \zsl \zlo} e^{\frac{i \bo^2}{2 r_{\rm F}^2}} \int \dd^2 \bb \, \psi (\bb) \, e^{- \frac{i \omega  \bb \cdot \bo}{\zlo}}\,,\label{eq:DiffIntegralBorn}
\end{align}
where we have expanded the time delay $t_d$, used Eq.~\eqref{eq:ExpansionB} to neglect the term $ \propto \bb^2$, and defined the Fresnel scale\footnote{Notice that the usual Fresnel scale is $r_{\rm F}'=\sqrt{ \zsl\zlo/(\zso \omega)}$, while here we have defined $r_{\rm F}$ as its projection onto the observer plane.}, $r_{\rm F}=\sqrt{\zso \zlo/(\zsl \omega)}$. Using Eq.~\eqref{eq:MomentumTransfer}, the plane wave in the expression~\eqref{eq:DiffIntegralBorn} can also be written as  $e^{- \frac{i \omega  \bb \cdot \bo}{\zlo}} = e^{- i \bb \cdot {\bm p}' } =  e^{- i \bb \cdot {\bm q} } $, if one promotes $\bb$ to a 3D vector with null $z-$component, since ${\bm p}$ and $\bb$ are orthogonal to each other. See~\cite{Yarimoto:2024uew} for a systematic treatment beyond the leading-order Born approximation.

At this point we are ready to compare the total wave computed within the lensing formalism to the results obtained with scattering techniques. As we will show, these two procedures will give the same result up to a relative difference in the overall normalization. 
In particular, by equating the full lensed and scattered fields $\phi_\omega (\bm r)$ of Eqs.~\eqref{eq:DefAmplificationFactor} and~\eqref{eq:ScatteringProblem}, we can write the matching condition as
\begin{align}\label{eq:MatchingBornTotal}
     \phi^0_{\omega, {\rm S}} + \frac{{\cal M}}{8 \pi \sqrt{s}} \frac{e^{i \omega r}}{r}  = {\cal N}  \llp \phi^0_{\omega, {\rm L}} + (F_{\rm diff} - 1 ) \phi^0_{\omega, {\rm L}} \rrp \,.
\end{align}
First, we fix the relative normalization ${\cal N}$ by matching the free waves close to the scattering surface ({\it i.e.} far away from the source, where the spherical wave has flat wavefronts). This yields~\cite{Pijnenburg:2024btj} 
\be 
{\cal N} \,  \phi^0_{\omega, {\rm L}} \approx {\cal N} \, \frac{e^{i \omega \zsl}}{\zsl} \, \phi^0_{\omega, {\rm S}}= \phi^0_{\omega, {\rm S}}  \quad \to \quad {\cal N} = \frac{\zsl}{e^{i \omega \zsl}}\,.
\ee 
Then, the matching of Eq.~\eqref{eq:MatchingBornTotal} implies
\begin{align}\label{eq:MatchingBornM}
   \frac{{\cal M}}{8 \pi \sqrt{s}} 
    = \frac{\zsl \zlo}{\zso} e^{-\frac{i\bo^2}{2 r_{\rm F}^2}} ( F_{\rm diff} - 1 ) \, ,
\end{align}
where we have used the large distance expansions to write $e^{i \omega r}/r  \approx {\rm exp} [i \omega (\zlo + \bo^2 / 2 \zlo)]/\zlo$ and $\phi^0_{\omega, {\rm L}} \approx {\rm exp} [i \omega (\zso + \bo^2/2 \zso)]/\zso$.
Using Eq.~\eqref{eq:DiffIntegralBorn}, we find
\begin{align}
  {\cal M} ({\bm p}, {\bm p}+{\bm q}) &= - \sqrt{s}  \int \dd^2 \bb \,e^{- i \bb \cdot {\bm q}}  \, \llp 4\omega^2 \psi (\bb) \rrp  \, .
\end{align}
Taking the Fourier transform definition in Eq.~\eqref{eq:VFT}, and the form of the two-dimensional gravitational potential in Eq.~\eqref{eq:2DPot}, the matching condition~\eqref{eq:MatchingBornM} then gives 
\begin{align}
    {\cal M} ({\bm p}, {\bm p}+{\bm q}) = \mathcal{V} ({\bm p}, {\bm p}  + {\bm q})\,,
\end{align}
which is the expected result of the scattering amplitude in the Born approximation, obtained as the lowest order solution of the Lippmann-Schwinger equation~\eqref{eq:LS}.

We can also perform this reasoning in the other direction, namely using Eq.~\eqref{eq:MatchingBornM} to infer $F_{\rm diff}$ in the Born regime, given the amplitude ${\cal M}$. For example, for a point mass lens, the gravitational Compton amplitude can be easily computed as the tree level $t$-channel scattering between a massless scalar (incoming wave) 
and a massive scalar (lens) interacting through a graviton, and is given by
\be
\mathcal{M} ({\bm p}, {\bm p}+{\bm q})=\frac{32 \pi G M^2 \omega^2}{|{\bm q|^2}} \,,
\ee
where the exchanged momentum ${\bm q}$ is related to the impact parameter as in Eq.~\eqref{eq:MomentumTransfer}.
Inverting Eq.~\eqref{eq:MatchingBornM}, we find that, in the Born approximation,
\be 
F_{\rm diff} - 1  =  \frac{r^2_{\rm E}}{\bl^2} \, {\rm exp} \llp \frac{i\nu }{2} \frac{\bl^2}{r_{\rm E}^2} \rrp\,,
\ee 
where we have introduced the dimensionless frequency $\nu =   4 G M \omega $, the Einstein radius $r^2_{\rm E} = 4 G M  \zsl \zlo / \zso $, and the impact parameter on the lens plane $\bl = (\zsl/\zso) \bo$. Our result matches the one in~\cite{Takahashi:2005sxa} in the limit $(r_{\rm E}/ |\bl|)^2\ll \nu$ or, equivalently, $|{\bm q}||\bl| \gg 1$. 

\vspace{0.1cm}
\noindent{{\bf{\em Partial waves and the Fresnel factor.}}} 
In the next two sections we will demonstrate that a similar relation between the scattering amplitude and the amplification factor  can also be derived in the context of a partial wave analysis. Indeed, it is known that, for central potentials, Eq.~\eqref{eq:HelmholtzLensing} can be solved by performing a decomposition over spherical harmonics. This is the standard lore in scattering theory, leading to the usual form of $f$ given in terms of the phase shifts $\delta_\ell$~\cite{Futterman:1988ni, Andersson:1995vi, Glampedakis:2001cx}
\be\label{eq:PWScattering}
f(\theta) = \; \sum^{\infty}_{\ell = 0} \frac{(2 \ell  +  1 )}{2 i \omega}  \lp e^{i \delta_\ell} - 1 \rp \, P_\ell (\cos \theta) \,,
\ee 
where the assumption of spherically symmetric potentials can be relaxed straightforwardly. Here, $P_\ell (\cos \theta)$ denotes the Legendre polynomial of the multiple moment $\ell$, in terms of the scattering angle $\theta$, and $\delta_\ell$ are the phase shifts encoding the information about the scattering. 
This form of the scattering amplitude is found by assuming an incoming plane wave in the $z-$direction, $\phi^0_{\omega, {\rm S}}  = e^{i \omega z}$, with partial wave expansion 
\begin{align}\label{eq:PWPlane}
\phi^{0}_{\omega, {\rm S}} ({\bm r})&=  \sum_{\ell = 0}^{\infty}  (2 \ell + 1) i^\ell j_\ell(\omega r) \, P_\ell (\cos \theta) \,,
\end{align}
where $j_\ell(\omega r)$ is the Bessel function. Because of the plane wave assumption, Eq.~\eqref{eq:PWScattering} does not strictly apply to a lensing scenario, where the incoming wave is taken to be spherical. Nevertheless, the type of incoming state does not change the physics of the scattering problem and the same phase shifts $\delta_\ell$ can be used to describe both scattering and lensing phenomena. To show that this is the case, one can perform a partial wave analysis when the incoming wave is spherical. Choosing boundary conditions such that $\phi^0_\omega (\bmr)$ is regular at $\bmr = 0$, and represents an outgoing wave as $r \to \infty$, the spherical wave $\phi^0_{\omega,{\rm L}} ({\bm r}) = e^{i\omega|{\bm r}-{\bm r}_{\rm s}|}/|{\bm r}-{\bm r}_{\rm s}|$ can be decomposed as
\begin{align}\label{eq:PWSpherical}
\phi^0_{\omega,{\rm L}}({\bm r})&=  i \omega\sum_{\ell = 0}^{\infty} (2 \ell + 1) (-1)^\ell j_\ell(\omega r) h^{(1)}_\ell (\omega r_{\rm s}) \, P_\ell (\cos \theta) \,,
\end{align}
where $r$ and $r_{\rm s}$ denote the relative distances between the lens and the observer or source, respectively, with $r < r_{\rm s}$,\footnote{Notice that choosing $r>r_s$ would demand swapping the arguments of the spherical Bessel and Hankel functions, yielding an equivalent result.} while the factor $(-1)^\ell$ arises from setting the source on the $-z$ axis. Comparing Eqs.~\eqref{eq:PWPlane} and~\eqref{eq:PWSpherical}, one can see that the only difference arises from the additional Hankel function $h^{(1)}_\ell (\omega r_{\rm s})$, if the incoming wave is spherical. 
Then, one can separate $j_\ell(\omega r)$ into its two spherical Hankel components, and collect the outgoing components of the full wave at large distances, to obtain~\cite{Nambu:2015aea,Motohashi:2021zyv}\footnote{The phase shifts $\delta_\ell$ are defined in terms of the asymptotic radial behavior of the  scalar field according to~\cite{Futterman:1988ni}
\begin{equation}
\phi_{\omega} ({\bm r}) \propto  r^{-1}e^{- i \omega r} + (-1)^{\ell+1} e^{2 i \delta_\ell} \, r^{-1} e^{i \omega r}\,, \qquad r \to \infty\,,
\end{equation}
when neglecting absorption processes.
}
\begin{align}
\label{phihh}
\phi_{\omega} ({\bm r}) - \phi^0_{\omega,{\rm L}}({\bm r}) & =  i \omega \sum_{\ell = 0}^{\infty} \frac{2 \ell + 1}{2}  (-1)^{\ell} \lp e^{i \delta_\ell} - 1 \rp \nonumber \\
& \times \llp h^{(1)}_\ell (\omega r)   h^{(1)}_\ell (\omega r_{\rm s}) \rrp  P_\ell (\cos \theta)\,.
\end{align}
Extracting the scattering amplitude $f$ from this expression requires taking the large distance limit, by approximating the spherical Hankel functions as $h^{(1)}_\ell (x) \xrightarrow[x\to \infty]{} (-i)^{\ell+1} e^{i x} / x$. At first order, one finds that the difference between an incoming plane or spherical wave thus reduces to an overall factor $e^{i \omega r_{\rm s}}/ \omega r_{\rm s}$ that can be reabsorbed into the normalization of the total wave.
However, since lensing results are typically obtained in the finite-distance limit, we include one additional order in the expansion of the spherical Hankel functions by considering~\cite{Turyshev:2021rkj,Turyshev:2018gjj} 
\be 
\label{h1Fresnel}
h^{(1)}_\ell (\omega r)  \approx  (-i)^{\ell +1}\frac{e^{i \omega r + \frac{i \ell(\ell + 1)}{2 \omega r}}}{ \omega r} \,.
\ee 
In other words, through the inclusion of this correction, the scattering amplitude retains a nontrivial radial dependence, in analogy to the diffraction integral, allowing for a matching prescription. Furthermore, let us stress that, as discussed in Ref.~\cite{Nambu:2015aea}, one could obtain the quadratic term of Eq.~\eqref{h1Fresnel} by considering a WKB solution to the wave equation. When following this approach, it is possible to improve the small $\ell$ behavior through the replacement $\ell(\ell+1) \to (\ell + 1/2)^2$, as shown by  Langer~\cite{Langer:1937qr}, getting the final expression
\be\label{eq:ScatteringWithFresnelLensing}
f(\bmr) =  \; \sum^{\infty}_{\ell = 0} \frac{(2 \ell  +  1 )}{2 i \omega} \lp e^{i \delta_\ell} - 1 \rp   e^{ \frac{i (\ell + 1/2)^2}{2 \omega } \llp \frac{1}{r} + \frac{1}{r_{\rm s }}\rrp }  \: P_\ell (\cos \theta) \,,
\ee 
which is the scattering amplitude for an incoming spherical wave with the inclusion of the quadratic {\it Fresnel} term.

\vspace{0.1cm}
\noindent{{\bf{\em Matching in the eikonal regime.}}} 
At this point, we are ready to demonstrate that the diffraction integral in Eq.~\eqref{eq:DiffractionIntegral} can be matched to the eikonal scattering amplitude obtained within the partial wave expansion.
To do so, let us consider the large $\ell$ limit of Eq.~\eqref{eq:ScatteringWithFresnelLensing}.  By defining the impact parameter on the lens plane as
\be
\label{eq:btol}
b = \frac{\ell + 1/2}{\omega}\,,
\ee 
the multipole sum can be approximated with a continuous integral $\sum_{\ell = 0}^{\infty} (2 \ell + 1 ) \approx 2 \omega^2 \int^{+\infty}_0 \dd b \, b$, while the Legendre polynomial simplifies to $P_\ell(\cos \theta) \approx  J_0 (\omega b \theta)$, in terms of the Bessel J function of order zero, $J_0(x)$, in the limit of small scattering angles $\theta \ll 1/\ell \ll 1$. The eikonal scattering amplitude then becomes 
\begin{align}\label{eq:ScatteringAmplitudeEikonal}
f^{\rm eik} (\bmr) &= - i \omega \int^{+\infty}_0 \dd b \, b \,   J_0 (\omega b \theta) \lp e^{ i \delta^{\rm eik}} -1 \rp e^{ \frac{i {b}^2 }{2r_{\rm F}^2} \frac{\zso^2}{\zsl^2} } \,,
\end{align}
in which $\delta^{\rm eik}$ is the eikonal phase shift, and we have approximated $r \approx z_{\rm lo}$ in the Fresnel factor.  While the derivation above has been provided for a spherically symmetric potential, it holds for arbitrary conservative potentials, in which case it becomes
\be
\label{eq:ScatteringEikonalReusm}
f^{\rm eik} (\bmr) =-i \frac{s-M^2}{4\pi \sqrt{s}} \int {\rm d}^2{\bm b}  \ e^{-i {\bm q}\cdot {\bm b}} \left(e^{i \delta^{\rm eik}}-1\right) e^{ \frac{i |{\bf b}|^2 }{2r_{\rm F}^2} \frac{\zso^2}{\zsl^2} } \,.
\ee
This is nothing but the standard result for the eikonal scattering between a massless particle\footnote{In the eikonal limit, the scalar field scattering amplitude reproduces the gravitational one, while subleading terms encode graviton helicity-flip effects (see comments in the Conclusions).} and a massive (lens) field, obtained by resumming all the ladder and cross-ladder diagrams in its Feynman series expansion~\cite{Abarbanel:1969ek,Levy:1969cr,tHooft:1987vrq},  with the addition of the finite distance Fresnel term. The phase shift in the eikonal limit is then given by \cite{Abarbanel:1969ek,Levy:1969cr,tHooft:1987vrq}
\begin{align}\label{eq:PhaseShiftEikonal}
    \delta^{\rm eik}(\bb) &=\frac{1}{2 (s-M^2)} \int  \frac{{\rm d}^2{\bm q}}{(2\pi)^2} \ e^{-i {\bm q}\cdot \bb} \mathcal{M}|_{G}({\bm{p}},{\bm p+ \bm q})  \nonumber  \\
    &= - 2 \omega \int \,  \dd z \,  U_{\rm L} \lp \bb,z \rp \,,
\end{align}
where $\mathcal{M}|_{G}({\bm q},s)$ describes the tree-level $t$-channel diagram. In the second line, we have used that $2 (s-M^2) \simeq 4 M \omega$, the leading order Lippmann-Schwinger equation~\eqref{eq:LS} to write the scattering amplitude in terms of the lens potential, and the Fourier transform conventions in Eq.~\eqref{eq:VFT}. 
Thus, comparing to Eq.~\eqref{eq:2DPot}, one derives
\be \label{eq:Match_eikonalphase_psi}
\delta^{\rm eik}(\bb) = -  \omega \psi(\bb)\,,
\ee
reproducing the known result that the two dimensional potential $\psi$ is equal to the gravitational time delay, $-\partial_\omega \delta$~\cite{Schneider:1992bmb}. In particular, for the  potential of a Newtonian pointlike lens, the eikonal phase shift takes the simple form
\begin{equation}
\delta^{\rm eik}(\bb) =- 4 G M \omega \log(| \bb |/ \xi)\,,
\end{equation}
in terms of an effective infrared scale $\xi$ of the lensing problem. 

Using the expression~\eqref{eq:Match_eikonalphase_psi} in Eq.~\eqref{eq:ScatteringEikonalReusm} then allows us to derive the matching condition in the eikonal limit
\begin{align}\label{eq:EikonalMatching}
    f^{\rm eik}  =  \frac{\zsl \zlo}{\zso} e^{-\frac{i \bo^2 }{2 r_{\rm F}^2 }  }  \lp F_{\rm diff} - 1 \rp \,,
\end{align}
which reproduces the matching relation~\eqref{eq:MatchingBornM} obtained in the Born approximation. 
This matching clarifies the crucial assumption of Eq.~\eqref{eq:DiffIntAssumptions-eik}, $|\partial_{\tilde r}^2 F| \ll |  \omega \partial_{\tilde r} F|$, which amounts to adopting the eikonal approximation $\ell \approx \omega b \gg 1$ [together with small deflection angles~\eqref{eq:DiffIntAssumptions-para}] as essential in drawing an analogy between the lensing diffraction integral and scattering frameworks, in addition to the condition of classicality listed in Eq.~\eqref{eq:ClassicalScattering}.

Finally, we note that the unitarity properties of the scattering amplitude allow one to derive optical-theorem–type statements for the diffraction integral (for work along these lines, but focusing on the consequences of causality, see~\cite{Tanaka:2023mvy,Tanaka:2025ntr,Suyama:2025gbh}). Since, by construction, the eikonal amplitude satisfies the optical theorem, the amplification factor given by the diffraction integral satisfies\footnote{Here we are assuming conservative potentials, but this can be easily generalized to include inelastic effects.}
\be
\frac{\zsl\zlo}{\zso}\int  {\rm d} \Omega \, |F_{\rm diff}-1|^2  = \frac{4\pi}{\omega} \, {\rm Im} [(F_{\rm diff}(\theta = 0)-1)] \ .
\ee
We note that this expression is not strictly valid in the Born regime, where the right-hand side vanishes. Additionally, for Coulomb potentials, one must regularize the divergences in the forward limit, $\theta=0$~\cite{Opt_Th_Coulomb_PhysRevLett.14.81,Review_Opt_Th_Coulomb}.


\vspace{0.1cm}
\noindent{{\bf{\em Post-Minkowksian corrections.}}} 
After drawing a precise analogy between the two frameworks, it is now possible to consider corrections to the leading eikonal scattering result and to add subleading corrections in the expression for the diffraction integral. In other words, the eikonal matching in Eq.~\eqref{eq:EikonalMatching} provides a way to go beyond the assumptions employed to derive it. 

A natural next step is to include corrections within the post-Minkowskian (PM) framework \cite{Bertotti:1956pxu, Kerr:1959zlt, Bertotti:1960wuq, Westpfahl:1979gu, Portilla:1979xx,Westpfahl:1985tsl}, which systematically accounts for weak-field relativistic corrections in scattering.  
In this setting, the lens can be systematically treated as a compact object with a well-defined gravitational potential given as a series expansion in $G$, allowing one to compute how scattering and lensing observables are modified beyond the Newtonian approximation.
To provide this PM extension, we first note that, while the
eikonal amplitude is obtained by resumming all the ladder and cross-ladder diagrams in its Feynman series expansion~\cite{Abarbanel:1969ek,Levy:1969cr,tHooft:1987vrq}, higher-order PM corrections to $\mathcal{M}^{\rm eik}$ arise from higher-order diagrams which are suppressed by powers of $ GM |{\bm q}|$. For instance, the leading correction to Newton, \emph{i.e.}, the 2PM correction, is captured by the classical part of the one-loop triangle diagram.

In the following, we will analyze the particularly clean example of a point mass lens, described by the Schwarzschild solution. At 2PM order, the correction is given by the classical contribution of the Compton amplitude in the eikonal limit, $\omega b \gg 1$, which, after using Eq.~\eqref{eq:LS}, gives the potential 
\begin{align}
\label{MPM}
    {\cal V}^{\rm eik}|_{G^2}(\bm{p}, \bm{p}+\bm{q}) &= \frac{32 \pi G M^2 \omega^2}{|\bm{q}|^2} + \frac{30 \pi^2 G^2 M^3 \omega^2}{|\bm{q}|}  \,.
\end{align}
This expression matches the Born series solution of a classical wave equation on a Schwarzschild background, as shown in Ref.~\cite{Correia:2024jgr}, and is in fact universal for non-spinning lenses.
In this case, it is possible to show that the scattering amplitude exponentiates again, obtaining an expression with the same structure of Eq.~\eqref{eq:ScatteringEikonalReusm}, where the corresponding phase shift is computed as 
\begin{align}
\label{delta2}
    \delta^{\rm eik}|_{G^2} (\bb) &= \frac{1}{4M \omega} \int  \frac{{\rm d}^2{\bm q}}{(2\pi)^2} \ e^{-i {\bm q}\cdot \bb} \, {\cal V}^{\rm eik}|_{G^2}({\bm{p}},{\bm  p+ \bm{q}}) \nonumber \\ 
   &= -4GM\omega \llp\log(| \bb |/ \xi) - \frac{15\pi}{16}  \frac{G M}{|\bb|} \rrp \,.
\end{align}
The second term in~\eqref{delta2} therefore represents the first nonvanishing PM correction to the eikonal phase shift for a Schwarzschild-like lens. As expected, this contribution is suppressed by powers of $GM/|\bb|$, and thus becomes relevant only when the lens lies sufficiently close to the emitting source.

To assess the relevance of this term in the diffraction integral, let us first introduce the dimensionless quantities  $ {\vec \varphi'} = \bb / (\zsl\theta_{\rm E})$ and ${\vec \varphi} = \bo / (\zso \theta_{\rm E}) $, expressed in terms of the Einstein angle $\theta_{\rm E} = \sqrt{4 M G {\zlo} / \zso \zsl}$.\footnote{Usually, the Einstein angle is defined as $\theta'_{\rm E} = \sqrt{4 GM \zsl / \zso \zlo}$. However,  this is the angle using the observer as center of the coordinate system. When using the source as center, then $\theta_{\rm E} = (z_{\rm lo}/ z_{\rm sl}) \theta'_{\rm E}$. } Then, the diffraction integral takes the form
\be \label{eq:FdiffPM}
F_{\rm diff}|_{G^2} =\frac{\nu}{2\pi i} \int {\rm d}^2{\vec \varphi'} e^{i\nu \left[ \frac12 ({\vec \varphi}'-{\vec \varphi})^2 - \log|{\vec \varphi}'| + \frac{{\cal C}|_{G^2} }{|{\vec \varphi}'|}\right]} \,, 
\ee
where
\be
{\cal C}|_{G^2} =  \frac{15 \pi}{64} \tilde{\theta}_{\rm E} \,, \qquad \tilde{\theta}_{\rm E} \equiv \sqrt{\frac{4  GM \zso}{\zsl \zlo}} \,,  
\ee
in terms of the Einstein ring measured from the lens position, $\tilde{\theta}_{\rm E} = (\zso / \zlo) \theta_{\rm E}$. This parameter dictates the strength of the PM correction and, as easily appreciated, takes larger values for heavier lenses or for GW sources sufficiently close to the lens itself. To provide few representative values, it is possible to show that, for characteristic LISA  binary systems located at  distances $\zso \simeq 100 \, {\rm Mpc}$, intermediate-mass black hole lenses $M \sim 10^3 M_\odot$, at distances $\zsl \simeq {\rm pc}$, would give ${\cal C}|_{G^2} \approx \mathcal{O}(10^{-5})$. On the other hand, one can consider hierarchical triples where the source orbits, at a distance $\zsl \simeq \mathcal{O}(10^{-3} \divisionsymbol 1) \, {\rm pc}$, a supermassive black hole lens with mass $M \sim \mathcal{O}(10^6 \divisionsymbol 10^8) M_\odot$~\cite{Antonini:2012ad,Berczik_2024}. In this case, the PM coefficient can reach values as large as ${\cal C}|_{G^2} \approx \mathcal{O}(10^{-2})$.

\begin{figure}
    \centering
    \includegraphics[width=\linewidth]{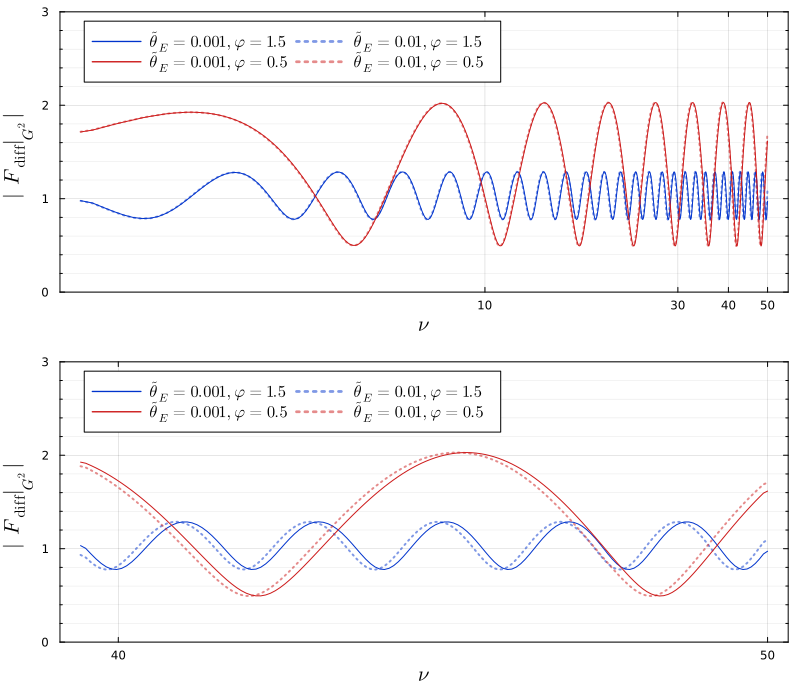}
    \caption{Behavior of the modulus of the diffraction integral in terms of the dimensionless frequency $\nu$, including the PM corrections at order $\mathcal{O}(G^2)$. Different colors show distinct values of the observer location $\varphi$ and the Einstein ring $\tilde{\theta}_{\rm E}$. The {\it bottom panel}  is a close-up  of  the {\it top panel} for  $\nu \in (40 , 50)$. }
    \label{fig:FPM}
\end{figure}

Following these estimates, we numerically evaluate Eq.~\eqref{eq:FdiffPM} to assess the effect of the PM correction on the diffraction integral. Figure~\ref{fig:FPM} illustrates the modulus of the diffraction integral as a function of frequency for two distinct choices of $\tilde{\theta}_{\rm E}$ (ranging from $10^{-3}$ to $10^{-2}$) and two values of the dimensionless observer position $\varphi = |\vec \varphi|$ (ranging from $0.5$ to $1.5$). The observed behavior can be explained as follows: in the low frequency regime ($\nu \ll 1$, see upper panel), the wave fails to resolve the source, resulting in a diffractive figure ($|F_{\rm diff}|_{G^2}| \to 1$) (the red curve in Fig.~\ref{fig:FPM} shows the same trend if one extrapolates it for smaller values of $\nu$). For higher frequencies, the formation of multiple images produces interference patterns, leading to the oscillatory behavior observed in $|F_{\rm diff}|_{G^2}|$.

\begin{figure*}
    \centering
\includegraphics[width=0.49\textwidth]{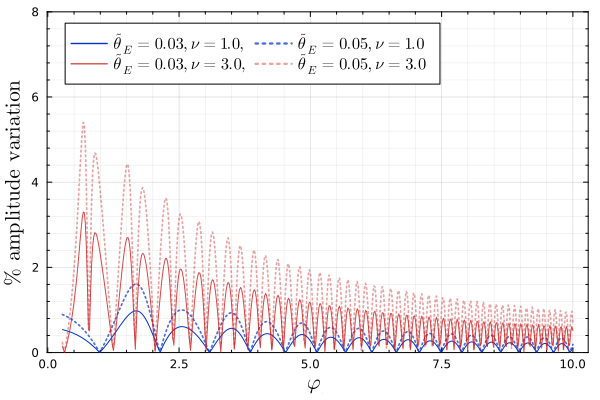}
\includegraphics[width=0.49\textwidth]{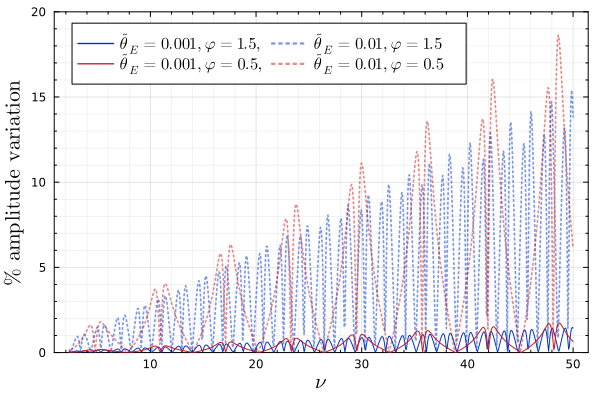}
    \caption{Percentage relative variation of the absolute value of the diffraction integral, accounting for the leading $\mathcal{O}(G^2)$ PM correction  to the phase shift.
    {\it Left panel}: Behavior in terms of the dimensionless impact parameter $\varphi$, changing the strength of the PM correction ${\cal C}|_{G^2} (\tilde{\theta}_{\rm E})$. {\it Right panel}:  Behavior in terms of the wave coefficient $\nu = 4 G M \omega$, for different values of the PM and impact parameters.}
\label{fig:beyond-PM}
\end{figure*}
\begin{figure*}[t!]
    \centering
\includegraphics[width=0.329\textwidth]{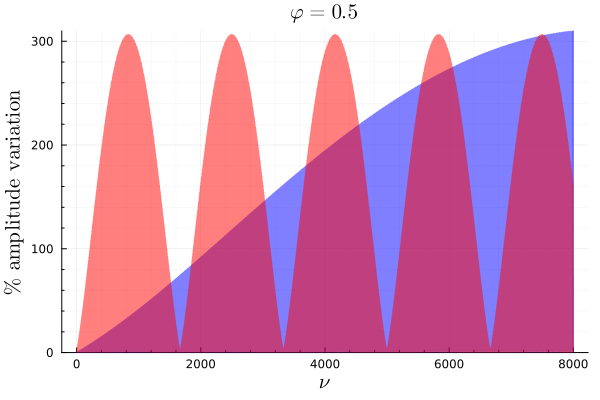}
\includegraphics[width=0.329\textwidth]{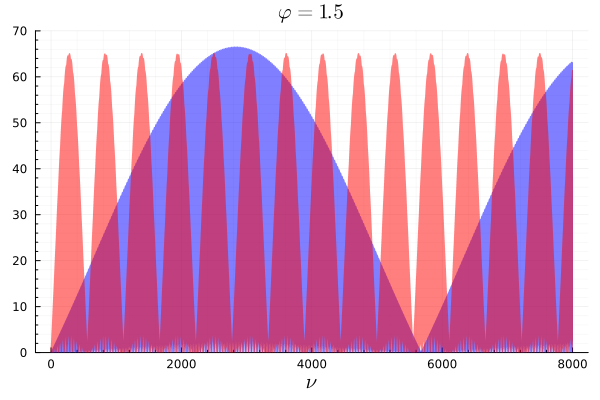}
\includegraphics[width=0.329\textwidth]{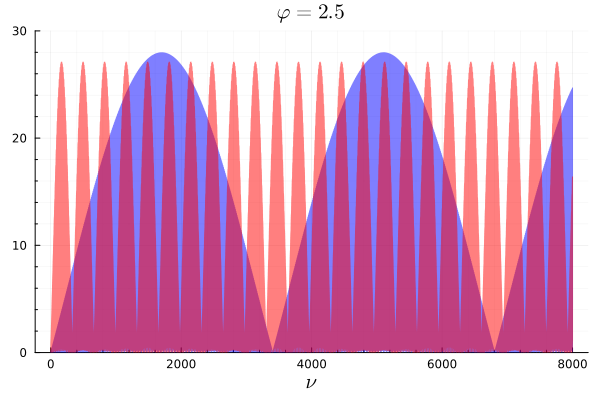}
    \caption{
    Same as right panel of Fig.~\ref{fig:beyond-PM}, but for larger values of the wave coefficient $\nu$. The {\it left}, {\it central} and {\it right panels}  correspond to the choices of the impact parameter $ \varphi = \{  0.5, 1.5, 2.5 \}$, respectively. Red and blue curves stand for different choices of the Einstein angle, namely $\tilde{\theta}_{\rm E} = \{ 0.01, 0.001 \}$ in order.}
\label{fig:beyond-PM-LargeNU}
\end{figure*}

Figures~\ref{fig:beyond-PM} and~\ref{fig:beyond-PM-LargeNU} show the percentage variation in the diffraction integral amplitude due to 2PM corrections relative to the 1PM-order (Newtonian) result of Eq.~\eqref{eq:DiffractionIntegral}, $||F_{\rm diff}|_{G^2}|/|F_{\rm diff}| - 1|$.

In the left panel of Fig.~\ref{fig:beyond-PM}, we show the percentage difference as a function of the transverse observer position at fixed dimensionless frequency $\nu$, illustrating that the impact of the PM corrections decreases with increasing $\varphi$. This behavior can be understood as follows: for large $\varphi$, paths with $\varphi' \ll 1$, which pass close to the lens, accumulate a larger geometrical and gravitational time delay, producing stronger phase differences between neighboring paths and leading to destructive interference. Consequently, the diffraction integral at large $\varphi$ is dominated by paths with $\varphi' \approx \varphi$, which stay farther from the lens and are therefore less affected by the PM correction.

The right panel of Fig.~\ref{fig:beyond-PM}, together with Fig.~\ref{fig:beyond-PM-LargeNU}, shows the frequency behavior of the fractional variation of the diffraction integral amplitude, at fixed observer position, and  can be understood as follows. The high-frequency oscillations in the ratio (see right panel of Fig.~\ref{fig:beyond-PM}) arise from the interference of the two geometric-optics images. The PM correction slightly shifts the relative phase between these images compared to the Newtonian case, as well as their magnifications, producing the fast wiggles observed in the diffraction pattern. As $\nu$ increases, the relative phase difference between the two images grows until it reaches $2\pi$, at which point the PM-corrected image phase difference realigns with the Newtonian ones. This produces a slow, long-wavelength modulation that acts as the envelope of the high-frequency interference (see Fig.~\ref{fig:beyond-PM-LargeNU}; the fast oscillations discussed before are present but not visible due to the density of the plot). The period of this modulation is given, at leading order in ${\cal C}_{G^2}$, by 
\begin{align}\label{eq:PeriodBeating}
    T_{\rm long} = \frac{2 \pi}{\Delta \Phi_{\rm PM}}\,, \quad 
    \Delta \Phi_{\rm PM} = {\cal C}_{G^2} \left( \frac{1}{\varphi_2} - \frac{1}{\varphi_1} \right)\,,
\end{align}
where $\varphi_{i}$ are the two images formed in a potential of a point-mass lens (without the PM-corrections), with the convention $\varphi_1 > \varphi_2$.\footnote{Notice that, when $\varphi_1 \approx \varphi_2$, the period of the modulation becomes very large, effectively suppressing the effects of the modification. This is the case when $\varphi = 0$, namely when source-lens-observer are all aligned.} 
As an example, we discuss the central panel of Fig.~\ref{fig:beyond-PM-LargeNU}. In this case, the observer position is set to $\varphi = 1.5$, and using the lens equation~\eqref{eq:b_lens_obs}, we find the position of the two images $\varphi_1 = 2.0$ and  $\varphi_2 = 0.5$, yielding a period of $T_{\rm long} \approx 5.7/ \tilde{\theta}_{\rm E} $.
This is in perfect agreement with Fig.~\ref{fig:beyond-PM-LargeNU}, where the red curve shows a beating pattern with frequency $\approx 570$, while the blue curve has a period of $\approx 5700$.  
Importantly, from Fig.~\ref{fig:beyond-PM-LargeNU} and Eq.~\eqref{eq:PeriodBeating}, we see that the value of $\tilde{\theta}_{\rm E}$ primarily fixes the frequency of the beating pattern (together with the image positions), rather than the amplitude of the percentage variation, which is mostly dictated by $\varphi$ (see the different panels of Fig.~\ref{fig:beyond-PM-LargeNU}). 
Indeed, at $\varphi = 1.5$, the correction reaches roughly $60\%$ (central panel of Fig.~\ref{fig:beyond-PM-LargeNU}), independently of $\tilde{\theta}_{\rm E}$.  

The conclusion is that PM corrections can produce sizable variations in the amplitude of the amplification factor when there is simultaneous destructive interference between the Newtonian images and constructive interference between their PM corrections or, vice versa, when the Newtonian images interfere constructively while their PM corrections interfere destructively.
Note that the PM correction also formally produces a third image; however, it is extremely faint, and its contribution is negligible, so it does not affect the description presented above.

\smallskip
Remarkably, the result of Eq.~\eqref{delta2} has been recently generalized  in~\cite{Bini:2025ltr,Bini:2025bll} (following inspiration from \cite{Ivanov:2025ozg}) by including all orders in the PM expansion, with the corresponding phase shift in the eikonal limit given by
\begin{align} \label{eq:resummed_delta}
    \delta^{\rm eik}(\bb) &= \pi |\bb|\omega-4GM\omega \log(| \bb |/ \xi) \nonumber\\
    &- \pi |\bb|\omega\, {}_3F_2\big[-\tfrac{1}{2},\tfrac{1}{6},\tfrac{5}{6};\tfrac{1}{2},1;27G^2M^2/\bb^2\big]\\
    &+\frac{64 }{3} \frac{G^3M^3\omega}{\bb^2} {}_4F_3\big[1,1,\tfrac{5}{3},\tfrac{7}{3};2,\tfrac{5}{2},\tfrac{5}{2};27G^2M^2/\bb^2\big]\,.
    \nonumber
\end{align}
Indeed, this result can be obtained as the  WKB solution to the radial wave equation in a Schwarzschild lens potential, and it is equal to the on-shell action for a null geodesic. 
While a detailed study of the effects of the PM-resummed eikonal phase in gravitational lensing is left for future work, we expect these corrections to become relevant at large deflection angles. In particular, while the PM expansion of Eq.~\eqref{delta2} cannot reproduce the multiple (relativistic) images that appear when the source is sufficiently compact, the fully resummed eikonal phase should, in principle, be able to do so. Indeed, computing the scattering angle 
\(\chi = -\frac{\partial \delta}{\partial |\bb|}\) 
shows that it diverges logarithmically as 
\(\chi(\bb) \sim \log(|\bb|-b_{\rm cr})\) 
at the critical impact parameter \(b_{\rm cr} = 3\sqrt{3}\,GM\), corresponding to the shadow of the lens. In this regime, \(\chi(\bb)\) becomes a multi-valued function with a branch cut at the shadow radius, leading to an infinite number of stationary points of the lens equation---each associated with a distinct image corresponding to a null geodesic that orbits the lens multiple times. By contrast, the PM-expanded scattering angle $\chi \sim 1/|\bb| + \mathcal{O}(1/|\bb|^2)$, derived from Eq.~\eqref{delta2}, yields only two such solutions (three when the 2PM correction is included), describing small-angle deflections. Thus, while the PM expansion captures the perturbative regime, only the fully resummed eikonal phase encodes the full image multiplicity. This strong field limit near the black hole shadow has been analyzed in \cite{Bozza:2001xd}, and a partial wave treatment was carried out numerically  in~\cite{Nambu:2015aea}, but Eq.~\eqref{eq:resummed_delta} can provide a more thorough analysis. We leave a detailed investigation to future work.


\vspace{0.1cm}
\noindent{{\bf{\em Beyond eikonal corrections.}}} 
The partial wave expansion in Eq.~\eqref{eq:ScatteringWithFresnelLensing} is exact in the far distance limit, and it was our starting point for deriving the scattering amplitude in the eikonal limit, showing that this matches the diffraction integral. We can improve this discussion by including beyond-eikonal (BE) corrections in the large $\ell$ limit of  Eq.~\eqref{eq:ScatteringWithFresnelLensing}, which correspond to finite-wavelength corrections of  ${\cal O} (\ell^{-1}) \sim {\cal O} (1/b\omega)$.
These corrections can arise from two sources: the large $\ell$ expansions of the phase shift $\delta^{\rm eik}$ and of the Legendre polynomials. The former can be extracted for a relativistic point-mass lens from the far-zone scattering phase shifts computed in black hole perturbation theory. For instance the phase shift for a spin-$s$ wave is given to 2PM order by  \cite{Casals:2015nja,Bautista:2023sdf,Ivanov:2024sds,Caron-Huot:2025tlq}
\begin{widetext}
\begin{align}
    \delta_\ell(\omega) &= -2GM\omega [\psi(\ell+1+s)+\psi(\ell+1-s)] +4GM\omega\log(\omega \xi) \nonumber \\
    &-\frac{4 \pi  G^2 M^2 \omega ^2}{2 \ell+1}\left[\frac{((\ell+1)^2-s^2)^2}{(2 \ell+1) (2 \ell+2) (2 \ell+3)} -\frac{(\ell^2-s^2)^2}{2 \ell (2 \ell-1) (2 \ell+1)}-\frac{s^2+2\ell
   (\ell+1)}{\ell
   (\ell+1)}\right] + {\cal O}(G^3) \,,
\end{align}
\end{widetext}
where here $\psi(x)$ is the digamma function (not to be confused with the lens potential above).
By taking the large $\ell$ limit, keeping $b$ as in Eq.~\eqref{eq:btol} fixed, we find
\begin{align}\label{eq:BECorrectionsSpin}
    \delta(\bb) \simeq &-4GM\omega \log(|\bb|/ \xi) + \frac{15\pi}{4} \frac{G^2 M^2\omega}{|\bb|} \\
    & -\frac{GM}{|\bb|^2\omega} \frac{1 - 12 s^2}{6} + \frac{G^2 M^2 \pi }{|\bb|^3 \omega } \frac{1 + 24 s^2}{16} + \cdots\,.\nonumber
\end{align}
The first line agrees with the eikonal phase shift in Eq.~\eqref{delta2} including its 2PM correction, and is independent of spin, in accordance with the equivalence principle\footnote{At the level of the partial-wave expansion, a further distinction between scalar and gravitational waves arises from the use of spin-weighted rather than scalar spherical harmonics, but this difference disappears in the eikonal limit for lenses which preserve helicity.}. The second line corresponds to sub-eikonal ${\cal O}(1/\ell^2)$ corrections, which depend on the spin of the wave. The correction at ${\cal O}(G)$ is purely Newtonian, whereas the second order one is a relativistic correction.

To simplify the discussion, we focus only on the first 2PM and beyond-eikonal corrections in~\eqref{eq:BECorrectionsSpin}, {\it i.e.}, we neglect the last term. For $|s| = 2$, the phase shift takes the form
\be
\delta^{\rm b-eik}|_{G^2}(\bb) = \delta^{\rm eik}|_{G^2} (\bb) + \frac{47}{6} \frac{GM \omega}{(|\bb|\omega)^2}\,.
\ee 
At this point, it is instructive to compare the magnitude of the two corrections, determined by the ratio of the characteristic scales $(GM/|\bb|)/(1/\ell) \sim GM\omega$. In the geometrical optics limit, $GM\omega \gg 1$, the PM corrections clearly dominate. Conversely, in the wave optics regime, $GM\omega \sim 1$, both corrections can yield significant contributions to the diffraction integral, with the beyond-eikonal term becoming dominant only in the deep wave limit.

To confirm such expectation, we can estimate the amplification factor according to the expression 
\be 
F_{\rm diff}^{\rm b-eik}|_{G^2} =\frac{\nu}{2\pi i} \int {\rm d}^2{\vec \varphi'} e^{i\nu \left[ \frac12 ({\vec \varphi}'-{\vec \varphi})^2 - \log|{\vec \varphi}'| + \frac{{\cal C}|_{G^2} }{|{\vec \varphi}'|} + \frac{{\cal C}|_{\rm b-eik} }{|{\vec \varphi}'|^2}  \right]} \,, 
\ee 
where we have now introduced a new coefficient capturing the beyond-eikonal corrections
\begin{align}\label{eq:CBE}
  {\cal C}|_{\rm b-eik} &= \frac{47}{24} \frac{\tilde{\theta}_{\rm E}^2}{\nu^2}\,.
\end{align}
Similarly to the discussion above, the size of this term is dictated by the size of the Einstein ring, together with the wave optics coefficient $\nu$. 
\begin{figure}
    \centering
    \includegraphics[width=\linewidth]{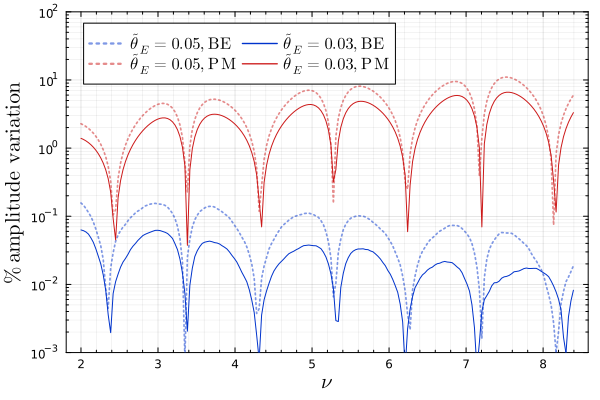}
    \caption{Percentage variation of the amplitude of the modulus of the amplification factor including only PM corrections (red lines) or only beyond-eikonal (BE) corrections (blue lines), as a function of $\nu$, for fixed $\varphi = 1.5$ and $\tilde{\theta}_E = \{ 0.03, 0.05\}$.  }
    \label{fig:beyond-eik}
\end{figure}
The impact of the beyond-eikonal corrections on the diffraction integral is shown in Fig.~\ref{fig:beyond-eik}, where we plot, for a fixed impact parameter $\varphi = 1.5$, the percentage corrections to the amplitude of the amplification factor when including either the PM correction (red curve, ${\cal C}|_{\rm b\text{-}eik} = 0$) or the beyond-eikonal correction (blue curve, ${\cal C}|_{G^2} = 0$) alone. 
In contrast to the PM corrections, the trend of the beyond-eikonal term is to decrease for larger values of~$\nu$, as expected from Eq.~\eqref{eq:CBE}. 
Therefore, one expects the dominant source of corrections at large~$\nu$ to arise from the PM terms, while at low~$\nu$ the beyond-eikonal corrections should, in principle, dominate. 
In practice, in our case, the beyond-eikonal contributions are always negligible compared to the PM corrections. 
This suppression is a consequence of the specific gravitational potential adopted here: for a Newtonian potential, the first beyond-eikonal correction vanishes~\cite{Bjerrum-Bohr:2019kec},\footnote{Let us mention that relaxing the assumption of spherical symmetry for the lens can in principle generate a nonvanishing leading beyond-eikonal correction, although such contributions are expected to be PM suppressed.} and the leading contribution scales as $\propto 1/\ell^2$, as shown in Eq.~\eqref{eq:BECorrectionsSpin}. 
Consequently, ${\cal C}_{G^2} \propto \tilde{\theta}_{\rm E}$, whereas ${\cal C}|_{\rm b\text{-}eik} \propto \tilde{\theta}_{\rm E}^2$, leading to the additional suppression seen in Fig.~\ref{fig:beyond-eik}.

An additional source of finite-wavelength corrections arises from the large-$\ell$ approximation of the partial waves. We show this by considering the partial-wave expansion of the outgoing wave in Eq.~\eqref{phihh} as,\footnote{To get this result, one has to consider the large distance and small angle limits, obtained by approximating the Hankel functions as 
\begin{align*}
    &h^{(1)}_\ell(\omega r) h^{(1)}_\ell (\omega r_{\rm s}) \approx (-1)^{\ell+1}\frac{e^{i \omega r} e^{\frac{i (\ell+1/2)^2}{2 \omega r}}}{ \omega r}  \frac{e^{i \omega \zsl}e^{\frac{i (\ell+1/2)^2}{2 \omega \zsl}}}{\omega\zsl} \nonumber \\
    &\approx (-1)^{\ell+1} \frac{\zso}{\omega\zsl\zlo}  e^{\frac{i \omega \bo^2}{2} \frac{\zsl}{\zso\zlo} } e^{\frac{i (\ell+1/2)^2}{2 \omega}\frac{\zso}{\zsl \zlo}} \frac{e^{i \omega |\bmr -\bmr_s|}}{\omega|\bmr -\bmr_s|}\,,
\end{align*}
where we have included the Fresnel factor as discussed below Eq.~\eqref{h1Fresnel}.
}
\begin{align}\label{eq:PhiOut}
&\phi_\omega({\bm r})
\supset\frac{e^{i \omega |\bmr -\bmr_s|}}{|\bmr -\bmr_s|}   \nonumber \\
&\times \Bigg\{ \frac{\zso e^{\frac{i  \bo^2}{2 r_{\rm F}^2}  }}{\zsl\zlo}   \sum_{\ell = 0}^{\infty} \frac{(2 \ell + 1)}{ 2 i \omega }  e^{ \frac{i (\ell + 1 /2)^2}{2 \omega } \frac{\zso}{\zsl \zlo}} \, e^{ i \delta_\ell} P_\ell (\cos \theta) \Bigg\}\,.
\end{align}
Using Hilb's asymptotic formula, the Legendre polynomial at large angular momenta can be written as~\cite{Amado:1985zz}
\be \label{eq:Hilb}
P_\ell(\cos \theta) \sim 
\sqrt{\frac{\theta}{\sin \theta}}
\left[
J_0 (\omega b \theta)
- \left( 1 - \theta \cot \theta \right)
\frac{J_1 (\omega b \theta)}{8\, \omega b \theta}
\right],
\ee
which is valid for small scattering angles, $\theta \ll 1/\ell$, with an error of order ${\cal O}(\ell^{-2})$, extending the approximation used to derive Eq.~\eqref{eq:ScatteringAmplitudeEikonal} to the leading sub-eikonal correction in $1/\ell$.
Looking at Eq.~\eqref{eq:PhiOut}, it is clear that the amplification factor is given by the parenthesis in the second line [see Eq.~\eqref{eq:DefAmplificationFactor}]. We then approximate the sum over $\ell$ as an integral over the impact parameter, improving the $P_\ell(\cos \theta) $ expansion as in Eq.~\eqref{eq:Hilb}, finding
\begin{align}
    F_{\rm diff}^{{\rm b-eik}, \, J_1} &= - i \omega \frac{\zso}{\zsl\zlo}  e^{\frac{i\bo^2}{2 r^2_F}  } \int^{+\infty}_0 {\rm d} b  \, b \,   e^{ \frac{i \omega b^2 }{2  } \frac{\zso}{\zsl \zlo}} \, e^{ i \delta }  \nonumber \\
    &\times \sqrt{\frac{\theta}{\sin \theta}} \llp J_0 (\omega b \theta) - \lp 1 - \theta \cot \theta\rp \frac{J_1 (\omega b \theta)}{8 \omega b \theta}  \rrp\,.
\end{align}
This integral can be computed analytically for  a Newtonian point mass lens with $\delta (b) = - \nu \log(b/\xi)$, yielding
\begin{align}\label{eq:BJ1Correction}
    &\frac{F_{\rm diff}^{{\rm b-eik}, \, J_1}}{F_{\rm diff}} -1= \sqrt{\frac{\theta}{\sin \theta}}  - 1 \nonumber\\
    &~~~~- \frac{1}{16}(1 - \theta \cot \theta) \sqrt{\frac{\theta}{\sin \theta}}  \frac{{}_1 F_1 \lp 1 + \frac{i\nu}{2} , 2 , \frac{i \theta^2}{2 \tilde{\theta}^2_{\rm F}} \rp }{{}_1 F_1 \lp \frac{i\nu}{2} , 1 , \frac{i \theta^2}{2 \tilde{\theta}^2_{\rm F}} \rp }\,,
\end{align}
where $\tilde{\theta}_{\rm F} \equiv \tilde{\theta}_{\rm E} / \sqrt{\nu} = \sqrt{\zso / (\omega \zsl \zlo)}$ is the Fresnel angle as measured from the lens position, outward toward the observer. 
Note that this expression is valid for $\theta \ll 1 / \ell$, i.e. $\theta \ll \tilde{\theta}_{\rm F}$, using $\ell \approx \omega b$ for large angular momenta.  
\begin{figure}
    \centering
\includegraphics[width=\linewidth]{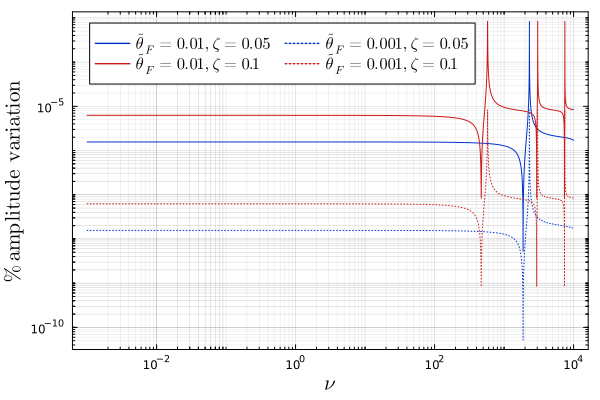}
    \caption{Percentage difference of the amplitude of the amplification factor due to the beyond-eikonal corrections, introduced by considering higher orders in the $P_\ell(\cos \theta)$ expansion, as a function of $\nu$ for different choices of $\tilde{\theta}_{\rm F}$ and $\zeta = \theta / \tilde{\theta}_{\rm F}$. }
    \label{fig:BEJ1}
\end{figure}
The percentage correction described by Eq.~\eqref{eq:BJ1Correction} is plotted in Fig.~\ref{fig:BEJ1} as a function of frequency for two choices of observer positions $\zeta = \theta / \tilde{\theta}_{\rm F} = \{ 0.05, 0.1\}$, and two choices of Fresnel scale $\tilde{\theta}_{\rm F} = \{0.001, 0.01 \}$. 
The plot shows that the $\mathcal{O}(1/\ell^2)$ correction is expected to give a small contribution to the full diffraction integral.

\vspace{0.1cm}

\noindent{{\bf{\em Discussion and conclusions.}}}
In this work, we have established a direct correspondence between the standard diffraction-integral formalism for GW lensing and the general theory of wave scattering. While the diffraction integral has long served as the central tool for modeling wave-optics effects, its physical meaning and the nature of the approximations that underlie it have remained largely implicit. By recasting gravitational lensing as a scattering problem, we have clarified these foundations and identified the precise regimes in which different approximations---Born, eikonal, partial-wave, and post-Minkowskian---remain valid.

We have shown explicitly that the diffraction integral is equivalent to the eikonal limit of the scattering amplitude for waves propagating in a weak gravitational potential, upon including finite-distance effects encoded in the Fresnel factor. This observation explains why the diffraction‐integral formalism smoothly interpolates between geometric and wave‐optics regimes and provides a clear organizing principle for systematic extensions. In particular, we have demonstrated how Born-like expansions, partial-wave techniques, and PM ({\it i.e.}, relativistic) corrections furnish controlled generalizations of the standard treatment. Within this unified framework, PM  corrections are especially relevant in the geometric-optics regime, where the wavelength is much smaller than the lens scale and higher-order terms refine the classical deflection, image structure, and time-delay predictions.
Conversely, beyond-eikonal corrections could become important in the wave-optics regime, where diffraction and interference dominate, and subleading corrections to the eikonal phase capture finite-wavelength and off-stationary-path effects. While in the case of a Newtonian point particle, these corrections are largely suppressed, we note that a leading beyond-eikonal correction can also be induced in theories beyond general relativity with screening mechanisms, in which the potential in the screened regions is no longer the usual Newtonian one \cite{Joyce:2014kja, Brax:2020ujo, Carrillo-Gonzalez:2021mqj}. Together, these extensions provide a systematic and improvable route to make accurate lensing predictions across all frequency regimes, yielding a compact and versatile framework that links GW lensing to well-developed scattering-amplitude techniques and enables controlled analytic modeling of lensing signatures in forthcoming GW observations.

\begin{figure}
    \centering
\includegraphics[width=\linewidth]{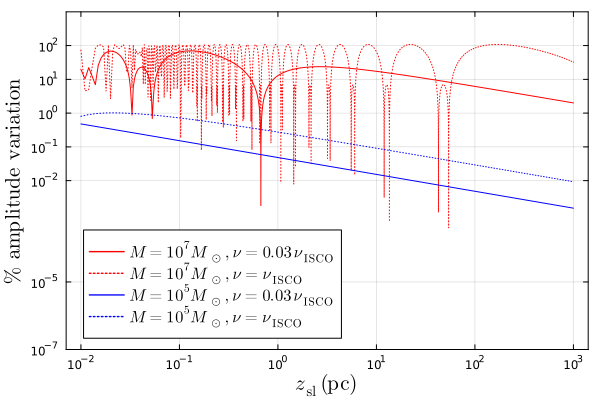}
    \caption{Percentage variation of the modulus of the amplification factor---accounting only for PM corrections---for a hierarchical triple consisting of a $M_{\rm S}=30M_\odot$ equal-mass binary orbiting a supermassive lens of mass $M$, shown as a function of their relative separation and evaluated for frequencies up to the binary’s innermost stable circular orbit, $\omega_{\rm ISCO} = (6\sqrt{6}G M_{\rm S})^{-1}$. In the plot, we have chosen $\varphi = 1.0$ for the observer transverse position.}
    \label{fig:pheno}
\end{figure}

As a phenomenological illustration of our results, Fig.~\ref{fig:pheno} shows the impact of PM corrections on the amplitude of the diffraction integral for a $30M_\odot$ equal-mass binary assembled in a hierarchical triple around a supermassive black hole. As expected, such configurations maximize these effects due to the small separation between the binary and the central object. The figure demonstrates that heavier lenses or more compact triples can generate very large corrections, which become increasingly significant in the high-frequency regime---approaching the binary’s innermost stable circular orbit---underscoring the importance of including these terms in data analyses of lensed GW events.

Looking forward, the scattering-based formulation naturally interfaces with effective field theories, offering a controlled avenue for extending current models. Indeed, an effective theory description of the process allows one to go beyond the point-particle approximation and incorporate the spin~\cite{Kubota:2023dlz,Kubota:2024zkv} and finite-size effects of realistic lenses, as well as to account for non-elastic phenomena such as wave absorption and polarization, which become particularly relevant when considering images close to black hole shadows at large frequencies (for a recent work along this direction see~\cite{Chan:2025wgz}). At the same time, the lensing perspective may allow for novel approaches to scattering problems: saddle-point structures, multiple-image formation, and multi-lens configurations provide concrete, physically realized examples of semiclassical propagation in nontrivial potentials, potentially informing broader developments in high-energy and gravitational scattering theory. We plan to investigate these avenues in future work. 

\vspace{0.1cm}
\noindent{{\bf{\em Acknowledgments.}}}
A.G. acknowledges Job Feldbrugge and Sunao Sugiyama for generously providing the code that enabled the numerical calculations presented in this work and for helpful discussions.  J.P.M. thanks Donato Bini for discussions and comments. We also thank Jose Diego for useful comments on the draft.
M.C.G is supported by the Imperial College Research Fellowship.
V.D.L. is supported by NSF Grants No.~AST-2307146, No.~PHY-2513337, No.~PHY-090003, and No.~PHY-20043, by NASA Grant No.~21-ATP21-0010, by John Templeton Foundation Grant No.~62840, by the Simons Foundation [MPS-SIP-00001698, Emanuele Berti], by the Simons Foundation International [SFI-MPS-BH-00012593-02], and by Italian Ministry of Foreign Affairs and International Cooperation Grant No.~PGR01167.
This work was carried out at the Advanced Research Computing at Hopkins (ARCH) core facility (\url{https://www.arch.jhu.edu/}), which is supported by the NSF Grant No.~OAC-1920103.
A.G. is supported by funds provided by the Center for Particle
Cosmology at the University of Pennsylvania.
The work of M.T. is supported in part by the DOE (HEP) Award DE-SC0013528. 

\bibliography{refs}

\end{document}